\documentclass[aps,prb,twocolumn]{revtex4}

\usepackage{amssymb}
\usepackage{graphicx}
\usepackage{dcolumn}
\usepackage{bm}

\def\be{\begin{equation}}
\def\ee{\end{equation}}
\def\bea{\begin{eqnarray}}
\def\eea{\end{eqnarray}}

\begin{document}
\title{Universality of phonon transport in surface-roughness dominated nanowires}
\author{P. Marko\v{s}}
\affiliation
{Department of Experimental Physics, Comenius University in  Bratislava, 842 28 Bratislava, Slovakia}
\author{K. A. Muttalib}
\affiliation
{Department of Physics, University of Florida, Gainesville, FL 32611-8440, USA}
\today
\begin{abstract}
We analyze, both theoretically and numerically, the temperature dependent thermal conductivity $\kappa$ of two-dimensional nanowires with surface roughness. Although each  sample is characterized by three independent parameters - the diameter (width) of the wire, the correlation length and strength of the surface corrugation - our theory predicts that there exists a universal regime where $\kappa$ is a function of a single combination of all three model parameters. Numerical simulations of propagation of acoustic phonons across thin wires confirm this universality and predict a $d^{1/2}$ dependence of $\kappa$ on the diameter $d$. 
\end{abstract}

\maketitle
 
\section{Introduction}

The challenge of designing a good thermoelectric device is, in part, to find a thermoelectric material that is simultaneously an``electron crystal and phonon glass'' \cite{rev1,rev2}, i.e. a thermoelectric material with a large electrical but a small thermal conductivity. This combination allows a large thermoelectric current without much heat dissipation, leading to a high efficiency. Significant efforts have gone into the art of nanoengineering novel materials having such properties \cite{rev3,rev4}. On the other hand it has been proposed recently that a device consisting of a number of parallel nanowires with an external gate voltage can be used \cite{mh} to exploit an interplay of the material parameters with the thermodynamic parameters in the non-linear transport regime  that can have both a large thermoelectric efficiency and a significant power output \cite{hmn}. While the nanowires are not necessarily `phonon-glasses', strong \textit{surface disorder} can suppress phonon transport significantly in Si nanowires with diameters  $d < 100$ nm, as demonstrated in recent experiments \cite{boukai,li,hochbaum,lim}. This is particularly important in the context of thermoelectric devices  since surface disorder is expected to suppress phonons more than electrons if the electron mean free path is much smaller than the diameter of the wire. 

The effect of surface-roughness on phonon transport in nanowires has been studied numerically  using Monte Carlo \cite{MC1,MC2} and Molecular Dynamics \cite{MD1,MD2,MD3,MD4}  simulations as well as models using wave scattering formalism \cite{WS1,WS2,WS3,WS4,WS5,WS6}. These studies show that the thermal conductivity in such cases can be much smaller than when the surface scattering is fully diffusive. Other theoretical models have considered only diffusive boundary scattering, together with various scattering mechanisms within the bulk \cite{galvin,mingo,kempa,nika}. 
In a recent work \cite{ma} it has been argued that the suppression of phonon transport in a surface-roughness dominated nanowire can be understood within a simple theoretical model that incorporates scattering of propagating phonons off localized phonons, where the localized phonons appear as a result of an exact mapping \cite{tesanovich} from a model with surface disorder to a model having a smooth surface with additional channel-mixing pseudo-interactions. The model with localized phonons has clear predictions about how the thermal conductivity depends on the various parameters that characterize the surface disorder, as well as on the parameters that characterize the localized phonons. However, these latter parameters are phenomenological, and have not been obtained from any microscopic considerations.

As is clear from experiments \cite{li,hochbaum,lim}, the effect of surface disorder depends crucially on the way the wire is prepared, by Electroless Etching (ELE) or Vapor-Liquid-Solid (VLS) techniques. Evidently, nanowires prepared with ELE have thermal conductivity $\kappa$ significantly smaller than those prepared with VLS. For VLS wires as the diameter $d$ is decreased, the low-temperature behavior of $\kappa$ apparently changes from a $T^3$ dependence for $d=115$ nm to a $T^2$ dependence for $d=37$ nm \cite{li}. The high-temperature behavior ($T>100$K) shows a downturn consistent with the importance of umklapp scattering in this regime. Theoretical models taking into account all significant bulk scattering mechanisms, changes in dispersion relations and a diffusive boundary have been used to fit the experimental results for VLS wires \cite{mingo2}. 
In contrast, the ELE wires seem to be qualitatively different. The low $T$ behavior of ELE wires seem to follow a $T^2$ dependence for all $115 \le d\le 50$ nm, and the high $T$ behavior do not show any downturn up to $T=300$K. This suggests that the ELE wires might be in a regime where the surface disorder dominates over all other bulk scattering mechanisms.

In this paper we consider phonon propagation in thin two-dimensional nanowires with surface disorder only, as a simple model for phonon transport in the surface-roughness dominated regime. We first analyze the theoretical model of Ref.~[\onlinecite{ma}] based on the scattering of propagating phonons off localized phonons and obtain
the thermal conductivity $\kappa$ for wires of diameter (width) $d$ for a fixed length $L \gg d$.  The surface disorder is characterized by an rms height $h$ of the roughness profile and a correlation length $l_c$. Although $\kappa$ in general depends on all of the parameters $d$, $h$ and $l_c$, we show analytically in the present work that 
there exists a \textit{universal regime} where $\kappa$ depends only on a single combination of all three parameters. This universal regime should be observable  in all surface-roughness dominated nanowires. While we show the existence of a single parameter by analyzing the simple model of  Ref.~[\onlinecite{ma}], the actual dependence on all three parameters can not be obtained analytically. We therefore perform numerical simulations on wires with appropriate surface disorder and obtain the scaling parameter by fitting to a universal curve of thermal conductivity as a function of temperature. We show that in this universal regime  the low-temperature dependence of $\kappa$ is $T^2$, and the high temperature behavior is independent of temperature. In addition, the diameter dependence of the thermal conductivity turns out to be approximately $d^{1/2}$. More generally, $\kappa(T)$ can be expressed in terms of a single parameter $\zeta=\sqrt{l_cd}/h$, which is consistent with the experimental results of Ref.~[\onlinecite{lim}].   The universality holds only in the diffusive regime, characterized by a $1/L$ dependence of the thermal conductance on the length of the wire.

All of the above properties are consistent with the ELE wires and inconsistent with the VLS wires; we therefore conclude that the ELE wires are indeed in the surface disorder dominated regime. More importantly, the universal scaling predicts that there are different possibilities of combining the parameters $d$, $h$ and $l_c$ to reach the same level of thermal conductivity, which might allow flexibility in designing a good thermoelectric device based on nanowires. 

\section{Theoretical model}

In the theoretical model of Ref.~[\onlinecite{ma}], the problem of phonons propagating in a disordered wire with surface roughness is mapped on to a problem of propagating phonons along a wire with smooth surface and additional interaction with localized phonons, the localized phonons having properties determined by the characteristics of the original model of surface disorder. We will consider a two-dimensional (2D) system, with length $L >> d$. We characterize the surface disorder by a Lorentzian power spectrum
\be
S(q; \Delta,l_c)=\frac{\Delta^2l_c}{1+q^2l_c^2}; \;\;\; \Delta\equiv \frac{h}{d}.
\ee
It then follows from Ref.~[\onlinecite{ma}] that the scattering rate of propagating phonons scattering off localized phonons in surface-roughness dominated nanowires should depend on the combined roughness parameter 
\be
R_0\equiv \frac{\Delta^2}{l_c}.
\ee
In addition, the existence of localized phonons suggest that the scattering rate should also depend on the parameters of the localized phonons, namely the widths $\Gamma _i$ and the frequencies $\Omega_i$.  
For simplicity, we will assume a fixed boundary condition at the surface which will allow us to compare directly with the numerical studies of section III. While this will in effect leave out some of the low-frequency surface modes \cite{WS6}, we will argue later that the experiments with ELE wires are consistent with the absence (or very low density) of low-frequency localized phonons.   
Thus we expect the localized phonons to be a discreet set and to have typically high frequencies, of the order of $\sqrt{k/M}$ where $k$ is the spring constant associated with the atoms in the material and $M$ is the typical cluster-mass that takes part in the localized vibrations. Suppose the smallest frequency is  $\Omega_1$. For simplicity we will also assume that for a given disorder, the widths of the relevant localized phonons are approximately the same, i.e. $\Gamma_i\approx \Gamma$. This is a reasonable approximation since the width largely depends on the effective barrier height and width that characterizes a given surface roughness. The contribution from the localized phonons to the scattering rate is roughly proportional to \cite{ma} 
 \bea
\frac{1}{2\tau(\omega)} &\propto & R_0 \sum_i\Gamma_i\left[\frac{1}{(\omega-\Omega_i)^2+\Gamma_i^2}-\frac{1}{(\omega+\Omega_i)^2+\Gamma_i^2}\right]\cr
 &\approx & R_0 \sum_i \frac{2\omega\Omega_i\Gamma}{[(\omega-\Omega_i)^2+\Gamma^2][(\omega+\Omega_i)^2+\Gamma^2]}.
\eea
In the small temperature regime almost all contribution to the thermal conductivity comes from the small frequency regime $\omega \ll \Omega_1$, where the scattering rate can be approximated as 
\bea
\frac{1}{2\tau(\omega)} \to R_0\frac{2\omega\Omega_1\Gamma}{[\Omega_1^2+\Gamma^2]^2} \approx R_0\frac{2\omega\Gamma}{\Omega_1^3}.
\eea
Here only the localized phonon with the smallest frequency contributes and we have  assumed $\Gamma \ll \Omega_1$.

In the opposite limit of large $\omega \ge \Omega_1$, the scattering rate will be dominated by the resonant scatterings from each localized phonon at $\omega=\Omega_i$. The total contribution from all the localized phonons will then be approximately 
\bea
\frac{1}{2\tau(\omega)}\approx R_0\sum_i\frac{1}{\Gamma_i}\approx R_0\frac{n_{loc}}{\Gamma}
\eea
 where $n_{loc}$ is the number of localized phonons within the propagating band.
 Thus the factor determining the disorder dependence of the scattering rate is expected to be
 \bea
 \frac{1}{2\tau(\omega)} &\propto & \left( \frac{\Delta^2}{l_c}\right) \frac{\omega\Gamma}{\Omega_1^3}, \;\;\; \omega \ll \Omega_1\cr
 &\propto & \left(\frac{\Delta^2}{l_c}\right) \frac{n_{loc}}{\Gamma}, \;\;\; \omega \ge \Omega_1.
\eea

The transmission function is proportional to the scattering time $\tau(\omega)$, the inverse of the scattering rate, multiplied by the propagating phonon velocities $v_Lv_R=\omega^2(2\omega_0^2-\omega^2)$ with the band edge at $\sqrt{2}\omega_0$. The thermal conductivity  in the diffusive regime then can be written as 
\bea
\kappa=\int_0^{\sqrt{2}\omega_0} d\omega\;\omega\; \tau(\omega)\;\omega^2(2\omega_0^2-\omega^2)\frac{\partial \eta}{\partial T}
\eea
 where $\eta$ is the Bose distribution function
$
\eta\equiv 1/(e^{\omega/T}-1)
$
 and we have chosen the Boltzmann constant $k_B=1$. The derivative has the limits
 \bea
 \frac{\partial \eta}{\partial T}=\frac{\omega}{T^2}\frac{e^{\omega/T}}{(e^{\omega/T}-1)^2}& \approx & \frac{\omega}{T^2}\frac{T^2}{\omega^2} \sim \frac{1}{\omega}, \;\;\; \omega \ll T \cr
 &\approx & \frac{\omega}{T^2} e^{-\omega/T}, \;\;\; \omega \gg T.
 \eea

\subsection{Low temperature regime}
 
 For $T<\Omega_1$, we can approximate the $\omega$-integral as follows:
 \bea
 \kappa &\approx & \int_0^T d\omega \; \omega\; \tau(\omega)\omega^2\;2\omega_0^2\;\frac{1}{\omega} \cr
 &+& \int_T^{\sqrt{2}\omega_0} d\omega \; \omega\; \tau(\omega)\omega^2(2\omega_0^2-\omega^2)\frac{\omega}{T^2}e^{-\omega/T}.
 \eea
 The second integral can be neglected due to the exponential. Using the small frequency expression for the scattering rate, we can then write
 \bea
 \kappa &\propto & \frac{l_c}{\Delta^2}\frac{\Omega_1^3}{\Gamma}2\omega_0^2\int_0^T d\omega\; \omega \cr
 & = & \frac{l_c}{\Delta^2}\frac{\Omega_1^3}{\Gamma}\;\omega_0^2 T^2, \;\;\; T \ll \Omega_1.
 \label{low-T}
 \eea
 Thus the thermal conductivity has a $T^2$ dependence in the low temperature regime for all values of the diameter and disorder parameters. This seems to be the case for all of the ELE wires in the regime $50 \le d \le 115$ nm but not true for the VLS wires where the $d=115$ nm wire has a $T^3$ dependence. 

\subsection{High temperature regime} 
 
 In the regime $T>\Omega_1$, the thermal conductivity can be approximated as 
 \bea
 \kappa &\approx & \int_0^{\Omega_1} d\omega \; \omega\; \tau(\omega)\omega^22\omega_0^2\frac{1}{\omega} \cr
& + & \int_{\Omega_1}^T d\omega \; \omega\; \tau(\omega)\omega^2(2\omega_0^2-\omega^2)\frac{1}{\omega} \cr
 &+& \int_T^{\sqrt{2}\omega_0} d\omega \; \omega\; \tau(\omega)\omega^2(2\omega_0^2-\omega^2)\frac{\omega}{T^2}e^{-\omega/T}.
 \eea
 Again we neglect the third integral due to the exponential term. 
 The second term is proportional to 
 $
  \frac{l_c}{\Delta^2}\frac{\Gamma}{n_{loc}}\int_{\Omega_1}^Td\omega \; \omega^2(2\omega_0^2-\omega^2).
 $
 This is much smaller compared to the first term since $\Gamma \ll \Omega_1$, so that $\kappa$ is approximately given by the first term which is independent of $T$,
 \bea
 \kappa \propto \frac{l_c}{\Delta^2}\frac{\Omega_1^3}{\Gamma}\; \omega_0^2\Omega_1^2, \;\;\; T\gg \Omega_1.
 \label{high-T}
 \eea
Thus the thermal conductivity saturates in the high temperature limit, the saturation value depending on the roughness parameters as well as the diameter of the wire. 
The ELE wires show this saturation for $T>200$K, the saturation value increasing with $d$. 
In contrast, the thermal conductivity of the VLS wires have maxima between $100$K and $200$K, and do not saturate for any of the diameters $115\ge d\ge 37$ nm up to $T=300$K. 
We note that the crossover temperature $\Omega_1$ is large for all of the ELE samples, which justifies our assumption of a fixed boundary condition. At the same time,
the Van-Hove singularity in 2D at $\omega_{VH}$ does not affect the thermal conductivity significantly if $\Omega_1 < \omega_{VH}$.

\subsection{Conjecture for universality}
 
We will assume that while the width of the localized phonons depends on the roughness parameters,  the smallest frequency $\Omega_1$  does not. This is consistent with our initial assumption that the frequency is largely dictated by the mass of the atoms taking part in the localized vibrations. Then for the entire range of temperature, the thermal conductivity is proportional to the parameter 
\bea
C(d,h,l_c)\equiv \left( \frac{l_c}{\Delta^2}\right) \frac{1}{\Gamma}.
\eea

It is not clear how to obtain the roughness parameter dependence of $\Gamma$, but we expect it to increase with increasing $d$ and increasing $l_c$, but decrease with increasing $h$. In the absence of a detailed microscopic theory, we propose the following as a conjecture:
\bea
\Gamma \propto \frac{d^{\alpha}l_c^{\gamma}}{h^{\mu}}
\label{conjecture}
\eea 
where $\alpha$, $\gamma$ and $\mu$ are all positive. Then
\bea
\kappa &\propto & C(d,h,l_c)\propto \frac{d^{2-\alpha}l_c^{1-\gamma}}{h^{2-\mu}}.
\label{scaling}
\eea
This suggests that for a given constant $C(d,h,l_c)$, all plots of thermal conductivity as a function of temperature for various choices of the three parameters $d$, $h$ and $l_c$ should fall on top of each other if the exponents $\alpha$, $\gamma$ and $\mu$ are known. This universality is a feature of phonon transport only in the surface-roughness dominated nanowires. Unfortunately it is not possible to determine the exponents from the present theoretical model. In the following section we will test the conjecture of universality numerically and obtain the exponents $\alpha\approx 3/2$,  $\gamma\approx 1/2$ and $ \mu\approx 1$  for a two-dimensional wire. Thus the thermal conductivity will be shown to be a universal function of the single parameter 
\be
\zeta\equiv \frac{l_c^{1/2}}{h}\;d^{1/2}.
\label{zeta}
\ee

\section{Numerical simulations}

\subsection{The model}

We will compare (\ref{scaling}) with numerical simulations.
For simplicity, we will assume a fixed boundary condition at the surface which was also assumed in Sec II. 
In addition we will only consider longitudinal phonons, since within our approximation, adding transverse phonons should not change either 
the temperature dependence or the eventual scaling properties of thermal conductivity.
In our model, the sample is represented by a square lattice with lattice constant $a=1$. 
For the atom located at site $xy$, the wave equation reads
\begin{equation}\label{neq:1}
\sqrt{\frac{m_{xy}}{k}}\omega^2 u_{xy} = u_{x+1y}+u_{x-1y}+u_{xy+1}+u_{xy-1}
\end{equation}
with the atomic mass $m_{xy}= m =1$ and spring constant $k=1$. The size of the lattice is $d\times L$. 
In numerical simulations, $d$ increases from $64$ to $256$ and the length of the system was chosen to be $L > 1000$. This corresponds to nanowires of width $12-50$ nm and length $> 200$ nm.

\begin{figure}[t]
\begin{center}
\includegraphics[width=0.42\textwidth]{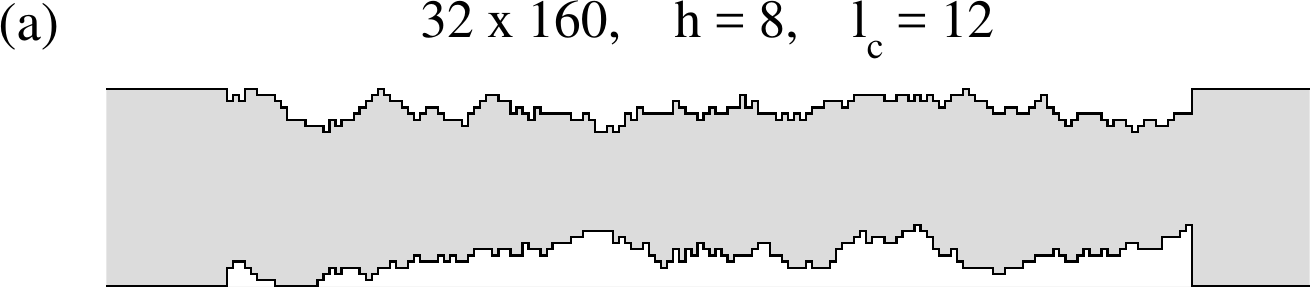}\\[5mm]
\includegraphics[width=0.20\textwidth]{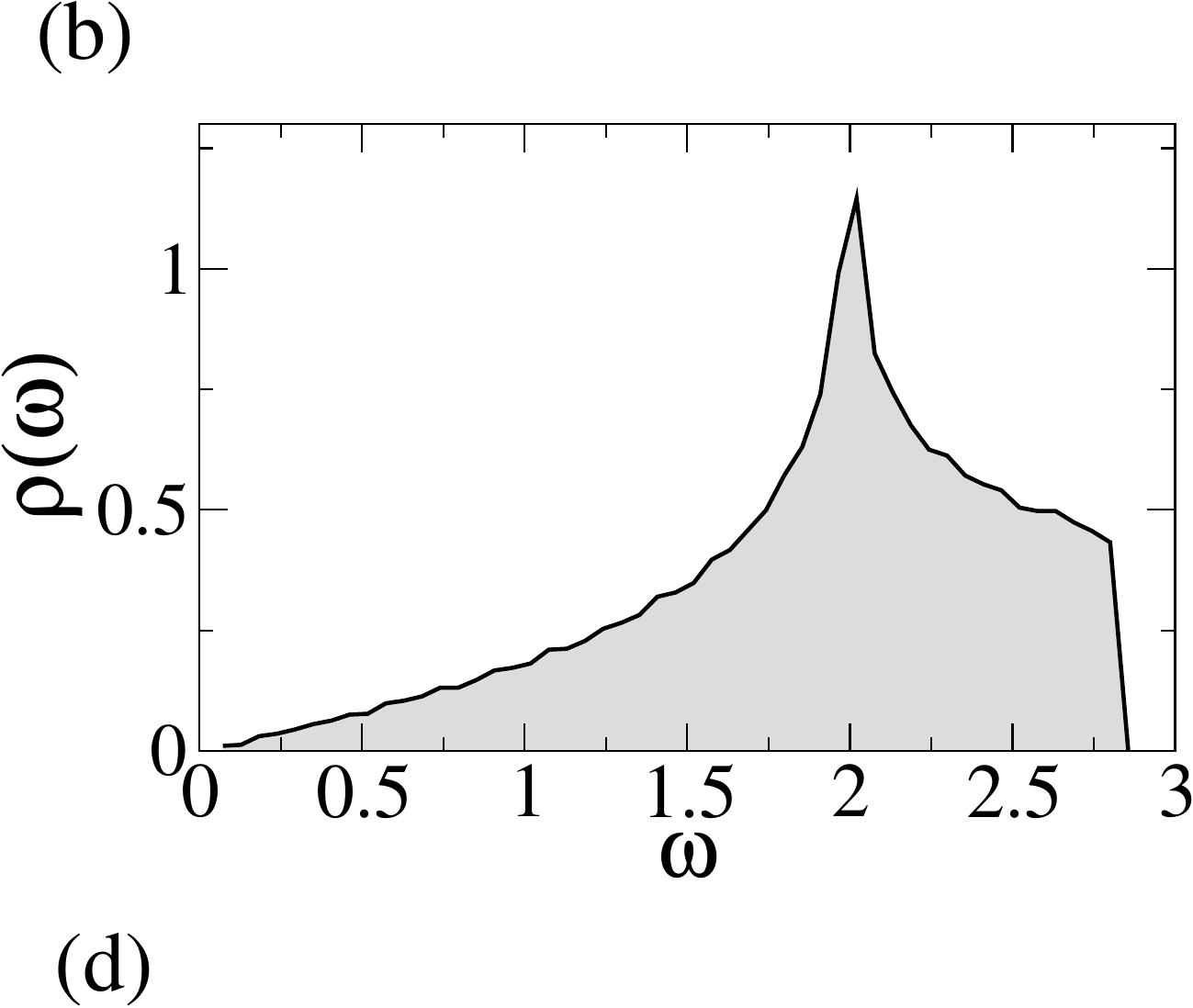}~~~~~\includegraphics[width=0.20\textwidth]{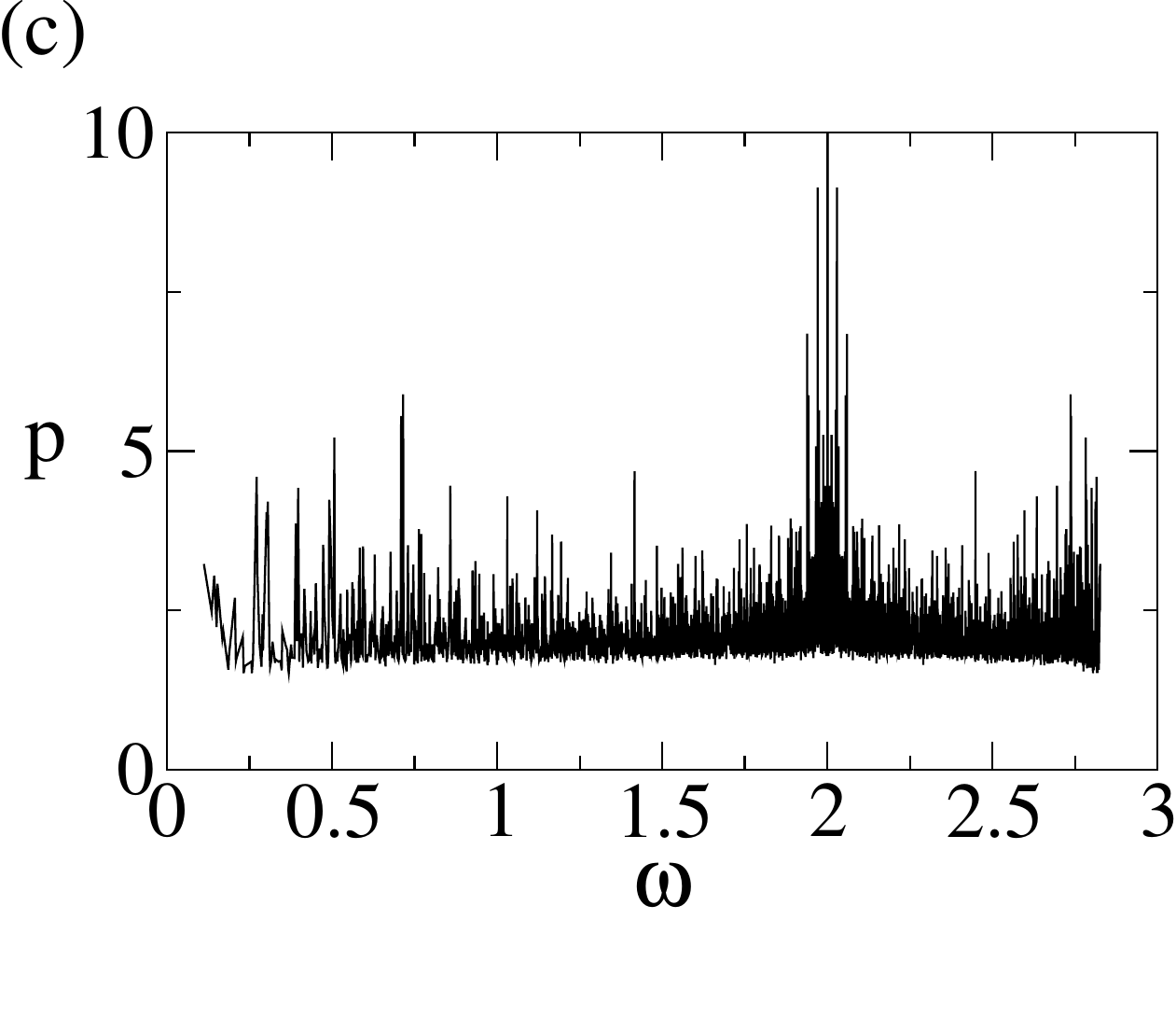}\\[2mm]
\includegraphics[width=0.42\textwidth]{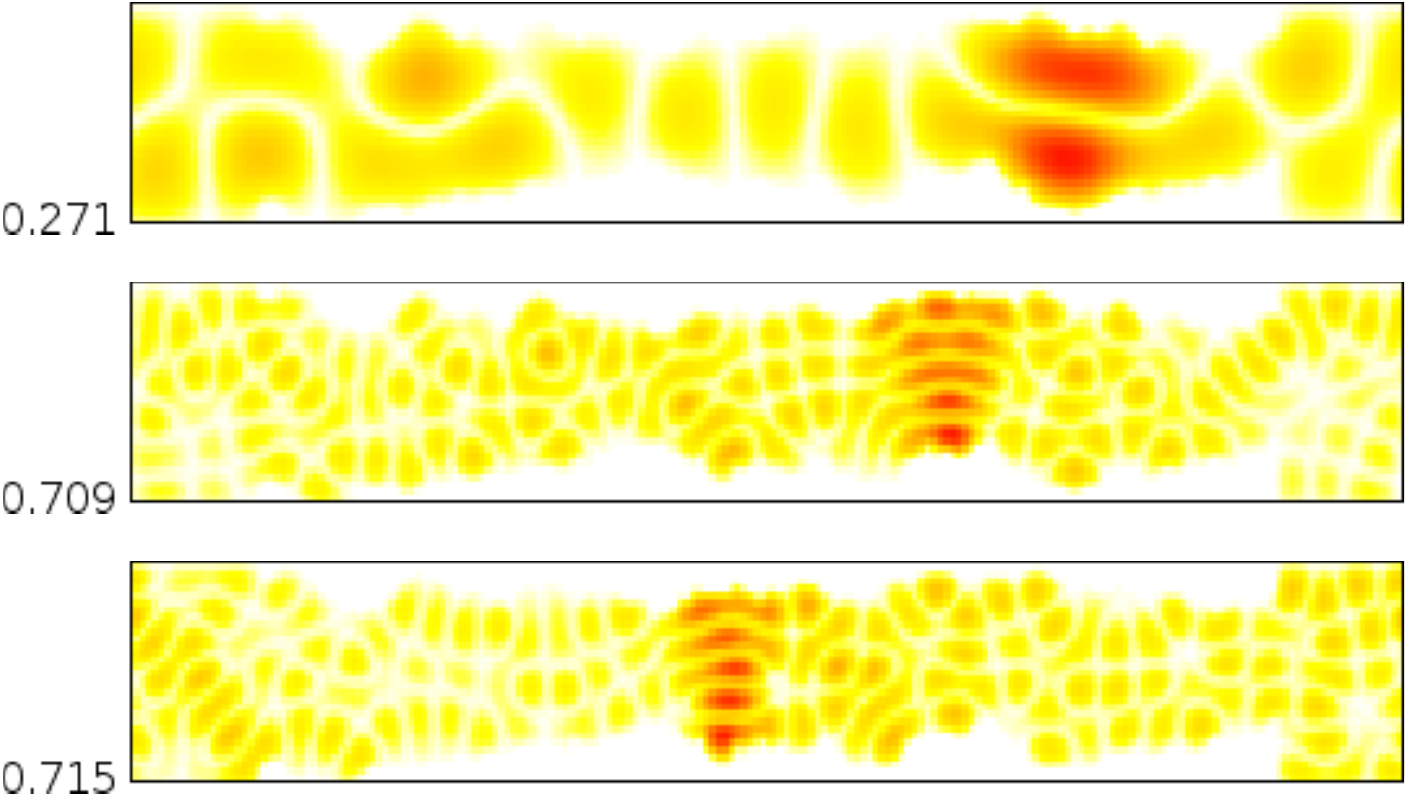}
\end{center}
\caption{%
(Color online) (a) Typical sample under study. Disordered sample is attached two two semi-infinite  leads.
(b) Density of states $\rho(\omega)$ for two-dimensional ideal square lattice.  The bandwidth is $2\sqrt{2}$. Note the van Hove singularity at $\omega=2$. (c) Inverse participation ratio $p$ as a function of frequency  calculated for sample (a) with periodic boundary conditions in horizontal direction. 
(d) Three  localized phonons with eigenfrequency $\omega = 0.27$, 0.709 and 
0.719. Shown is absolute value $|u_n(\vec{r})|$.  In red sites $|u_n(\vec{r})|>0.05$ with maximal values 0.1. White sites with zero phonon amplitude lie outside the wire. 
}
\label{fig-p}
\end{figure}

In order to create the surface disorder of the nanowire with appropriate $h$ and $l_c$,
we first generate a set of random numbers $\{\xi_x\}$, $x=1,2,\dots L$  with zero mean and correlation
$\langle \xi_x\xi_{x'}\rangle = h^2\exp -|x-x'|/l_c$. Then we  define a  surface profile  $y_x=\xi_x+\delta$ with constant shift $\delta =  - \textrm{min}~ \xi_x$ 
which guarantees that $y_x\ge 0$ for each $x$. 
Then, for a given $x$, we  substitute all atoms with $y\le y_x$ 
by heavy atoms with  mass $M=10^4m$.
The opposite  boundary of the sample is constructed in a similar way.
This restricts phonons to propagate only in the region occupied by ``light'' atoms. 
The sample is attached to two semi-infinite ideal leads of width $d$.
Figure \ref{fig-p}(a)  shows a typical sample. 

Thanks to spatial periodicity of the lattice, the frequency spectrum in the leads consists of one frequency band,
$0\le \omega \le 2\sqrt{2}$  with van Hove singularity, typical for 2D systems,  at $\omega=2$.
The phonon  density of states is shown in Fig. \ref{fig-p}(b). 

To check the presence of localized states in the sample, we calculate all eigenfrequencies  $\omega_n$ and  (normalized) eigenfunctions $u_n(\vec{r})$ of the structure shown in Fig. \ref{fig-p}(a). Localized states could be identified by analysis of the  inverse participation ratio \cite{MacKK}
\begin{equation}
p_n = \sum_{\vec{r}}|u_n(\vec{r})|^4 / \sum_{\vec{r}} |u^{(0)}(\vec{r})|^4
\end{equation}
where $u^{(0)}$ is (any) eigenfunction of the same system without surface disorder.
The eigenstate is localized if $p_n\gg 1$. 
The plot of  $p$ as a function of frequency shown in
Fig. \ref{fig-p}(c) confirms that localized phonons exist mostly in the vicinity of van Hove singularity, and their density is small in the low frequency part of the spectra.
Three  localized phonons are shown in Fig. \ref{fig-p}(d). 

\subsection{Thermal conductance}

We now use standard Economou-Soukoulis formula \cite{ES}
\begin{equation}
g = \sum_i T_i
\label{eq:land}
\end{equation}
to obtain the transmission $g= g(\omega)$ as a function of frequency. In Eq. (\ref{eq:land}), $g(\omega)$ is given as a sum of contributions 
$T_i$  of all open transmission channels.
Detailed analysis showed that $g(\omega)$ typically consists of contribution from ballistic transmission channels with $T_i \approx 1$, diffusive channels, as well as some localized channels with negligible transmission $T_i$.

In numerical simulations, we 
map the  wave model given by Eq. (\ref{neq:1}) into an electronic model
\begin{equation}
E\Phi_{xy} =
\Phi_{x+1y}+\Phi_{x-1y}+\Phi_{xy+1}+\Phi_{xy-1} + V_{xy}\Phi_{xy}
\end{equation}
where energy $E=\omega^2$ and potential $V_{xy} = (m_{xy}-1)\omega^2$.
The method is described in Refs.~[\onlinecite{MS1,MS2}]. 

\begin{figure}[t!]
\begin{center}
\includegraphics[width=0.22\textwidth]{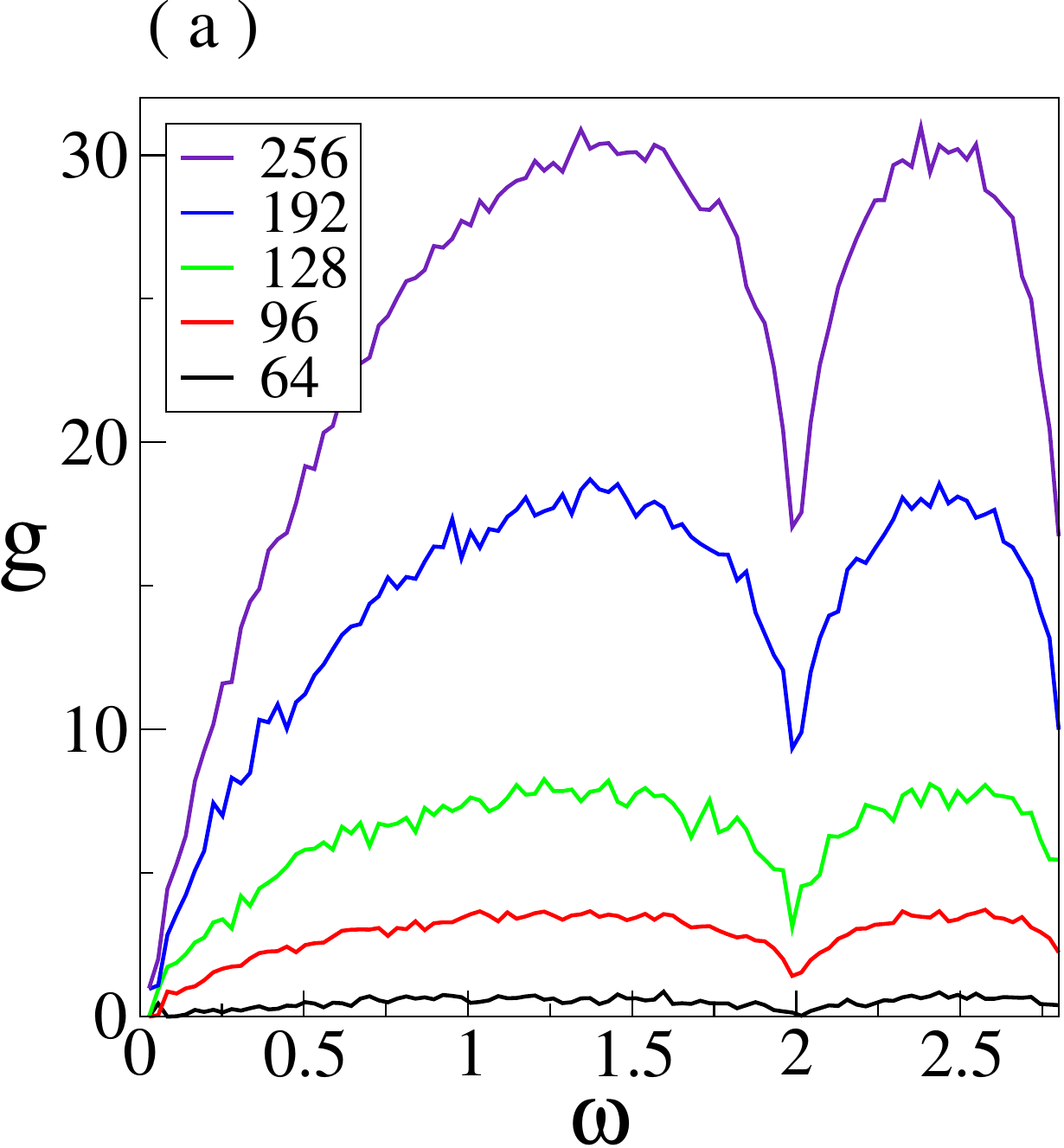}
~
\includegraphics[width=0.22\textwidth]{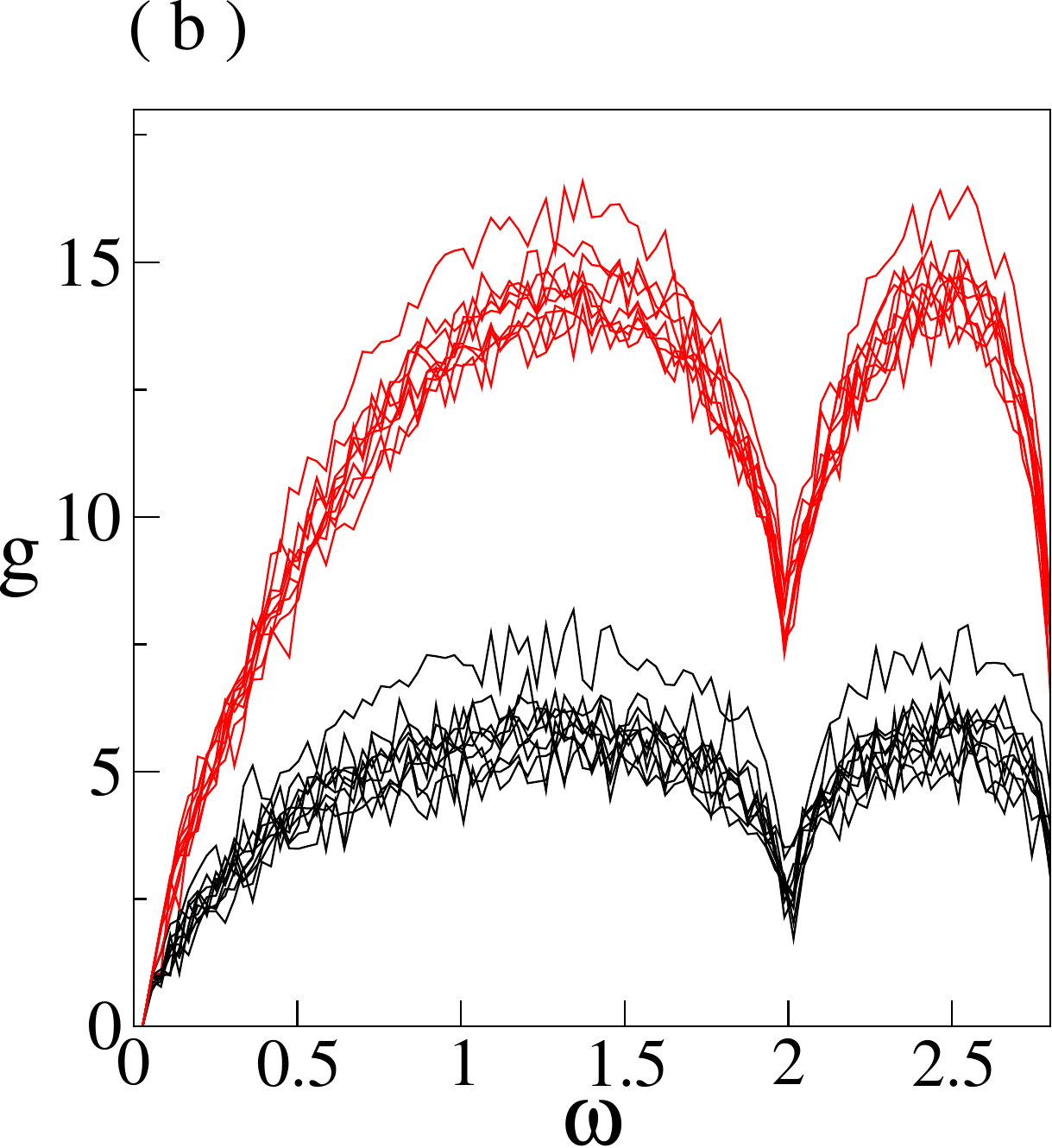}
\end{center}
\caption{%
(Color online) (a) $\omega$-dependence of $g$ 
for  samples of size $d\times 2048$, with $d = 64, 96, 128, 192$ and $256$. The surface roughness  $h=20$ and the correlation length $l_c = 100$.
(b) $g(\omega)$ for ten samples with different realization of surface disorder  and size $128\times 2048$, correlation length $l_c=200$. The surface disorder is $h=17$ (upper (red) curves) and $h=29.5$ (lower (black) curves).
}
\label{fig-A}
\end{figure}

Figure \ref{fig-A}(a) shows a typical frequency dependence of the transmission $g(\omega)$. 
For small frequency, the transmission increases linearly as $ \omega$. The dip in the transmission at $\omega=2$
corresponds to the van Hove singularity in the density of states.
Figure \ref{fig-A}(b) proves, in agreement with previous numerical studies \cite{saenz1,saenz2,mosko}, that the value of the transmission  depends on the realization of surface disorder. 
Observed  transmission  fluctuations  are of order unity in the diffusive transport regime and increase when  disorder increases.

\begin{figure}[t]
\begin{center}
\includegraphics[width=0.35\textwidth]{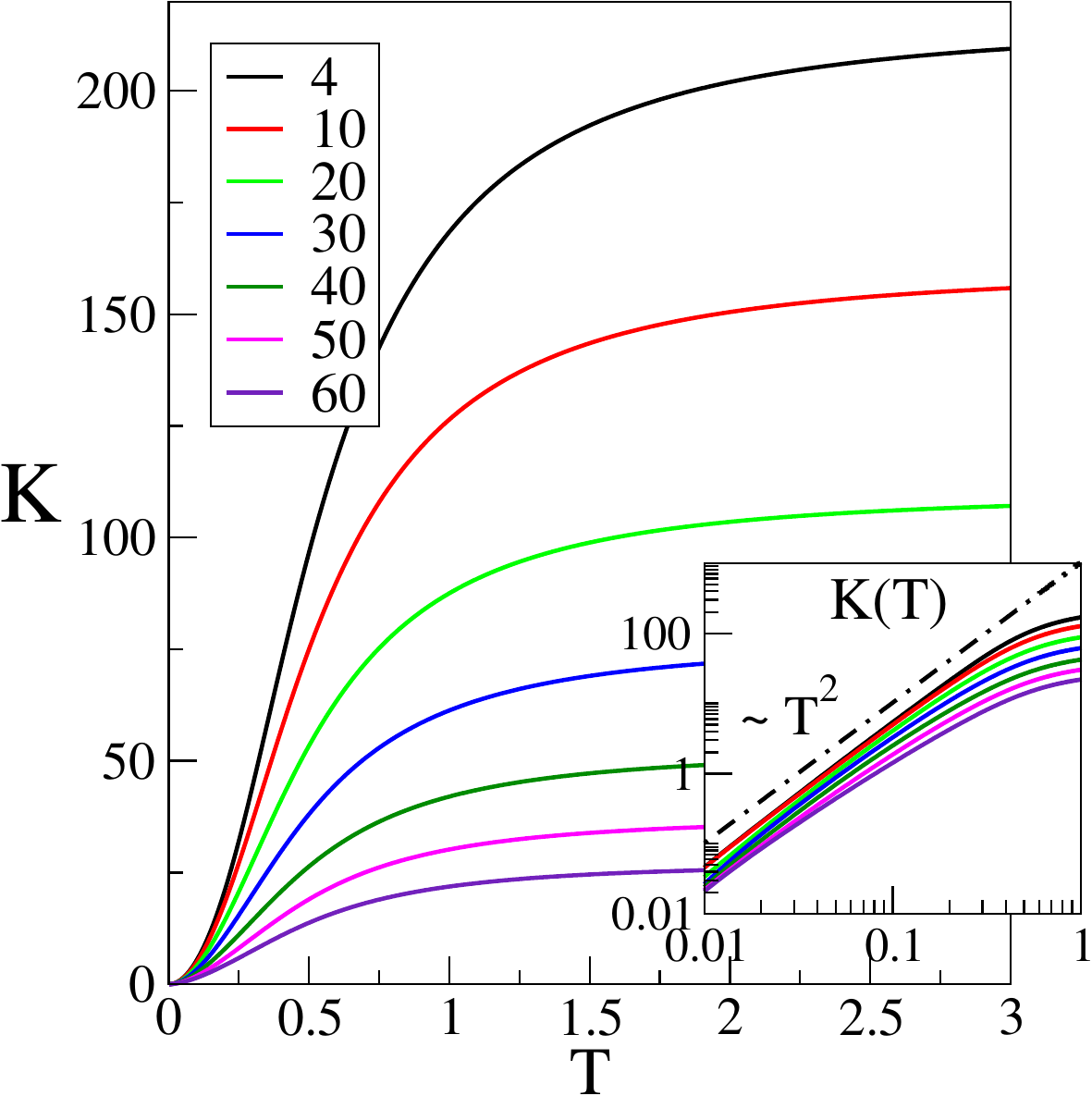}
\end{center}
\caption{%
(Color online)
Thermal conductance  $K$ given by Eq. (\ref{eq-kappa-1}) calculated numerically for 
sample size $d\times L = 256\times 2048$. 
Correlation length  $l_c=400$.
The surface roughness increases from $h=4$ to $h=60$. For all $h$, $K$ exhibits the same $T$-dependence:
Inset shows the $K(T)\sim T^2$ for small $T$. For large $T$,
$K$ saturates since our model does not include umklapp processes. 
}
\label{fig-BX}
\end{figure}

Numerical data provide us with the transmission $g(\omega)$ which determines the \textit{thermal conductance}  $K$  as
\begin{equation}
\label{eq-kappa-1}
K = \int_0^{\omega_D} \textrm{d}\omega g(\omega) \left[\frac{\omega/2T}{\sinh(\omega/2T)}\right]^2.
\end{equation}
The upper limit of the integration, $\omega_D$, lies above the upper edge of the frequency band. Typical temperature dependence of $K$ is  shown in Fig. \ref{fig-BX}, consistent with theoretical expectations of Eqs. (\ref{low-T}) and (\ref{high-T}).

Note that the thermal conductance does not coincide with thermal conductivity discussed in the previous section. $K$ depends on the geometrical size of the sample and is defined 
not only for the diffusive regime, but also for ballistic and localized regimes. In the diffusive regime and two-dimensional geometry, the conductivity $\kappa$  can be obtained as\cite{pichard}
\begin{equation}
\kappa = K\frac{L}{d}. 
\label{calK}
\end{equation}

Besides the phonon wavelength, our model introduces four length scales. Size of the sample is determined by its width $d$ and length $L\gg d$. The disorder is given by the strength of the surface roughness $h$ and correlation length $l_c$. Now we investigate how the thermal conductance depends on all of these parameters.

\subsection{Universality}

\begin{figure}[t!]
\begin{center}
\includegraphics[width=0.36\textwidth]{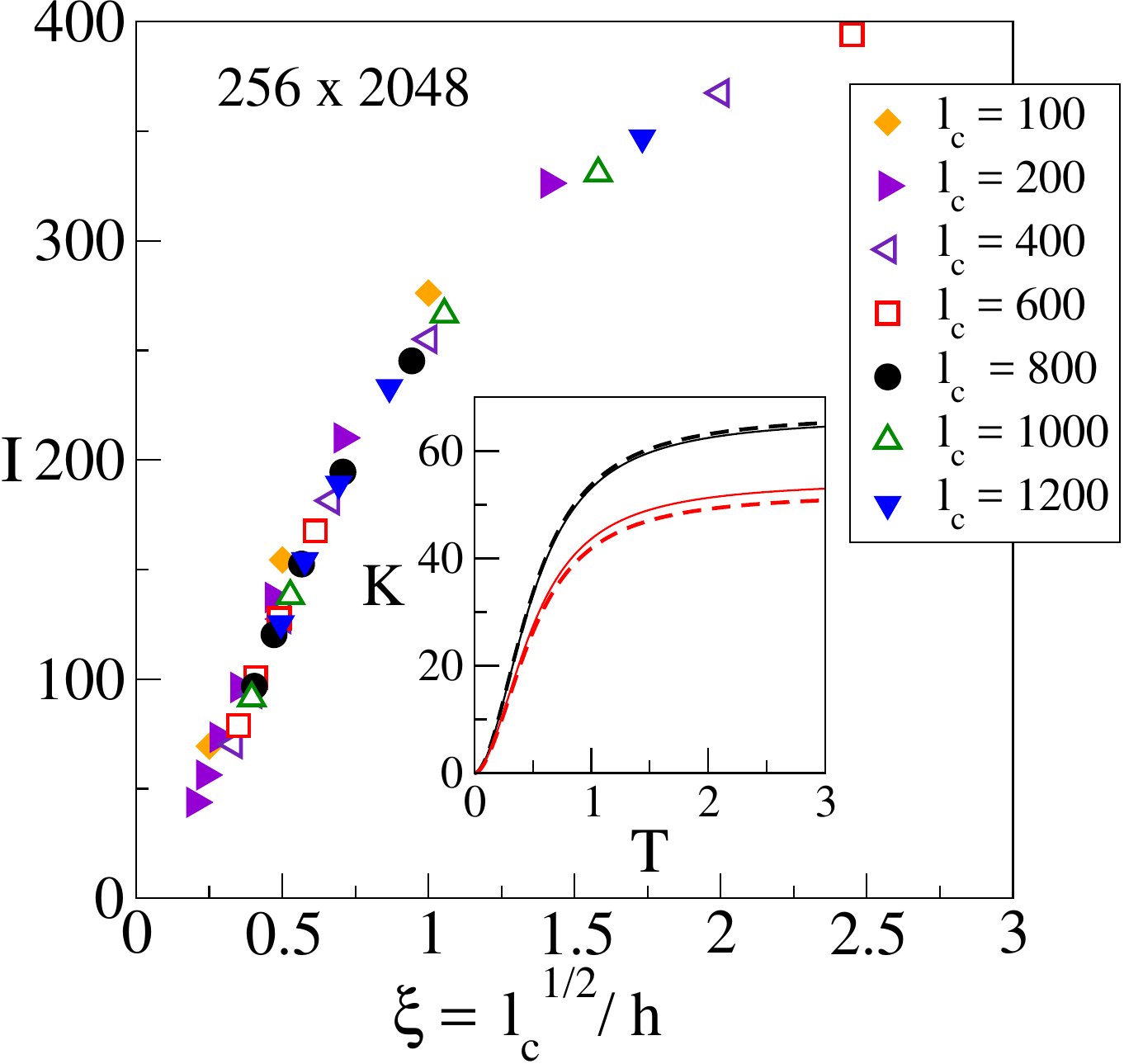}
\end{center}
\caption{%
(Color online)
Integral $I$, given by Eq. (\ref{eq-kappa-2}) for samples of size $256\times 2048$ and various values of $h$ and $l_c$. Data prove that for a given size of the system,  $I$ is a function of only one parameter, $\xi = \sqrt{l_c}/h$.
Inset shows the universality of the function $K(T)$ for samples with the same value of integral $I$: Black lines: 
$l_c=1200$, $h=60$ ($I= 154$) and $l_c=800$, $h=50$ ($I=152.5$).
Small deviation between two  displayed curves is given by different values of $\xi$.
Red lines  shows $K(T)$ two samples with 
$l_c=1200$, $h=70$ ($I= 125$) and $l_c=800$, $h=60$ ($I=120.3$).
}
\label{fig-ADU}
\end{figure}

To compare the thermal conductance for different samples, we introduce an integral
\begin{equation}
I = \int_0^{T_{\rm max}} \textrm{d}~T~K(T),
\label{eq-kappa-2}
\end{equation}
with the upper limit $T_{\rm max}=3$, and calculate how $I$ depends on the  model parameters. Owing to similar monotonic $T$-dependence of $K(T)$ (Fig.~\ref{fig-BX}), we expect that universality of $I$ guarantees the universality of $K(T)$ for any value of $T$. As an example, we show in the inset of Fig.~\ref{fig-ADU} the universality of conductance $K(T)$ for similar values of $I$. Furthermore, combination of Eqs. (\ref{eq-kappa-1}) and (\ref{eq-kappa-2}) guarantees
the universality of  $g(\omega)$   for each frequency $\omega$.

We find numerically that for a given system width $d$, $I$ depends only on the combination 
\begin{equation}
\xi = \frac{\sqrt{l_c}}{h}.
\end{equation} 
As an example, we show in Fig.~\ref{fig-ADU} the integral $I$ for system size $d\times L = 256 \times 2048$. Although the correlation length $l_c$ increases by an order of magnitude from $100$ to $1200$ and disorder $h$ varies between $4$ and $70$, all data collapse on a single curve. Similar universal $\sqrt{l_c}/h$-dependence  was obtained for other widths of the sample (not shown).

Inset of Fig. \ref{fig-ADU} confirms our assumption, namely that two samples with the same value of integral $I$ possess the same $T$ - dependence of the thermal conductance $K(T)$. 

\begin{figure}[t!]
\begin{center}
\includegraphics[width=0.42\textwidth]{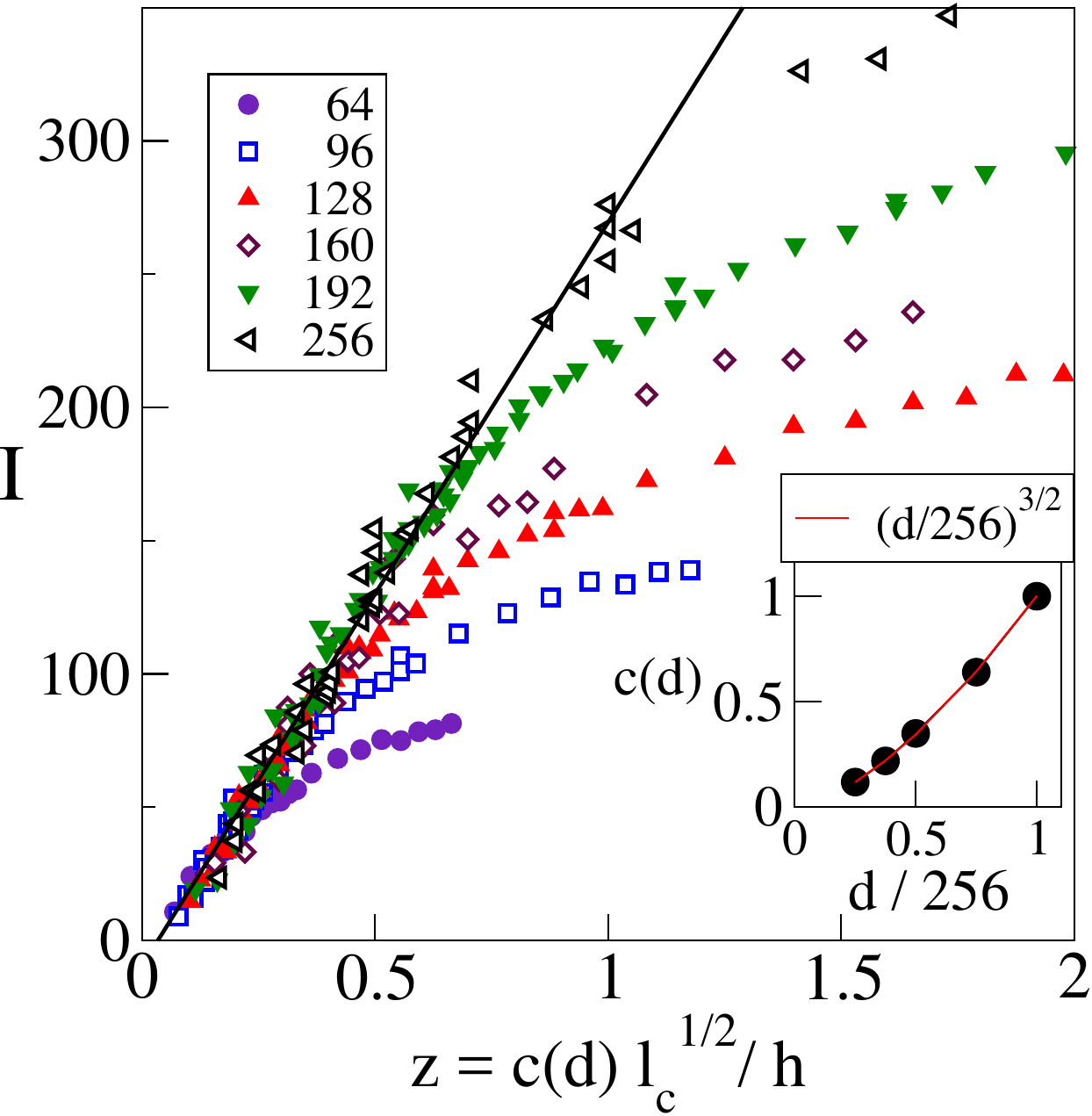}
\end{center}
\caption{%
(Color online)
Integral $I(l_c,h,d)$ for six widths of the system (given in legend) as a function of  parameter $z=d^{3/2}l_c^{1/2}/h$. The length of all systems is $L = 2048$.  
Solid line is a linear fit for $d=256$ for data $I<300$. The function $c(d)\propto d^{3/2}$ (shown in the inset) 
has been found to guarantee the overlap of data for all $d$ in the linear (diffusive) regime $I(z) \propto z$. 
}
\label{fig-ADD}
\end{figure}

Combining the results for different $d$, in Fig. \ref{fig-ADD} we plot $I$ as a function of the parameter 
$z=d^{3/2}\xi= d^{3/2} l_c^{1/2}/h$. Interval of $z$ in which $I(z)\propto z$ corresponds to the diffusive regime. 
As shown in Fig. \ref{fig-ADD}, the slope of this linear dependence is universal. 
Using Eq.~(\ref{calK}), the thermal conductivity $\kappa$ then has the universal dependence on the parameter $zL/d=\zeta L$ where $\zeta$ is given in Eq.~(\ref{zeta}). 
For larger values of
$z$, the system is in the ballistic transmission regime, and  the universality is lost. Similarly, for very small $z$ we expect to reach a non-universal localized regime.

\begin{figure}[t!]
\includegraphics[width=0.23\textwidth]{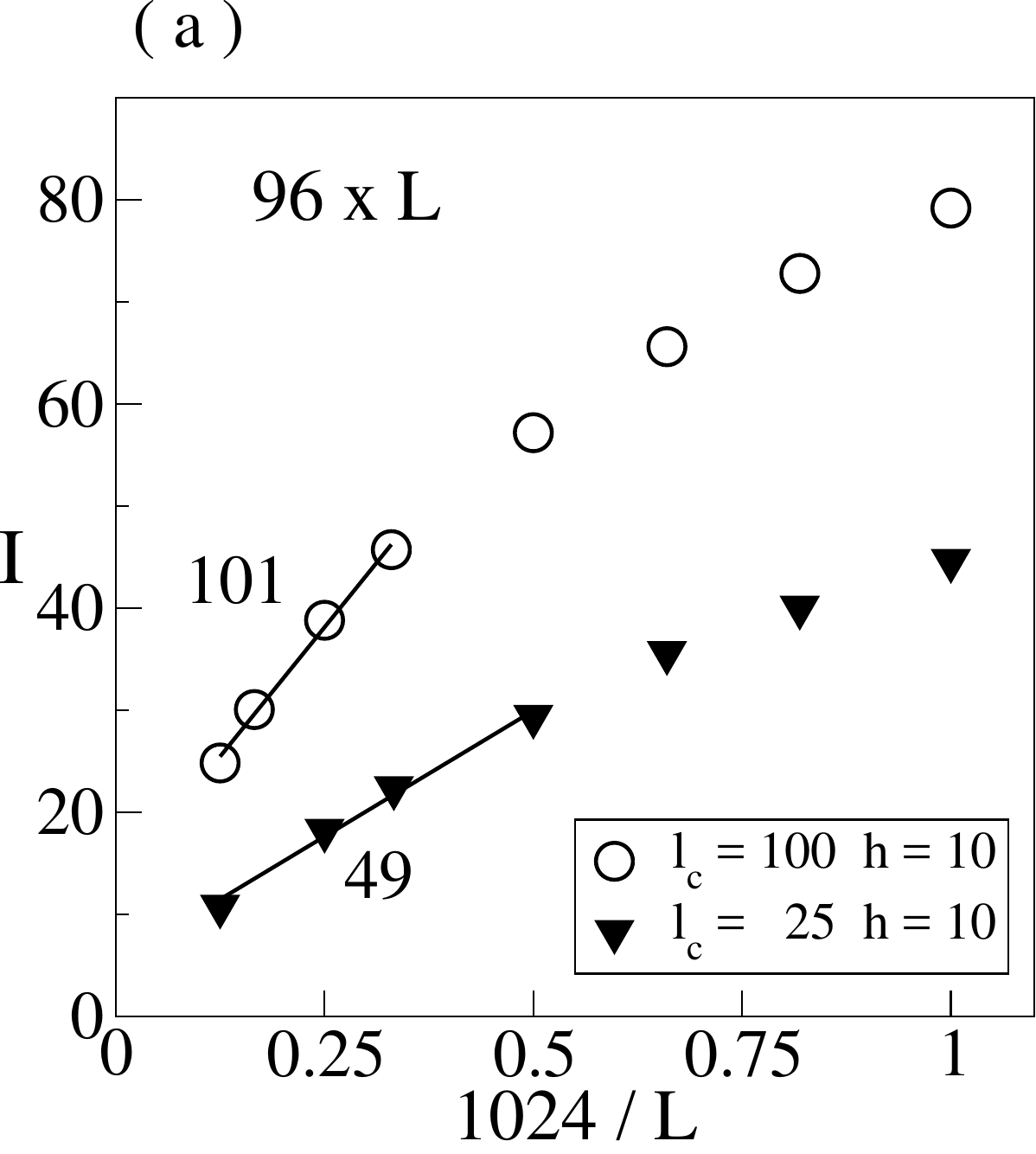}
\includegraphics[width=0.23\textwidth]{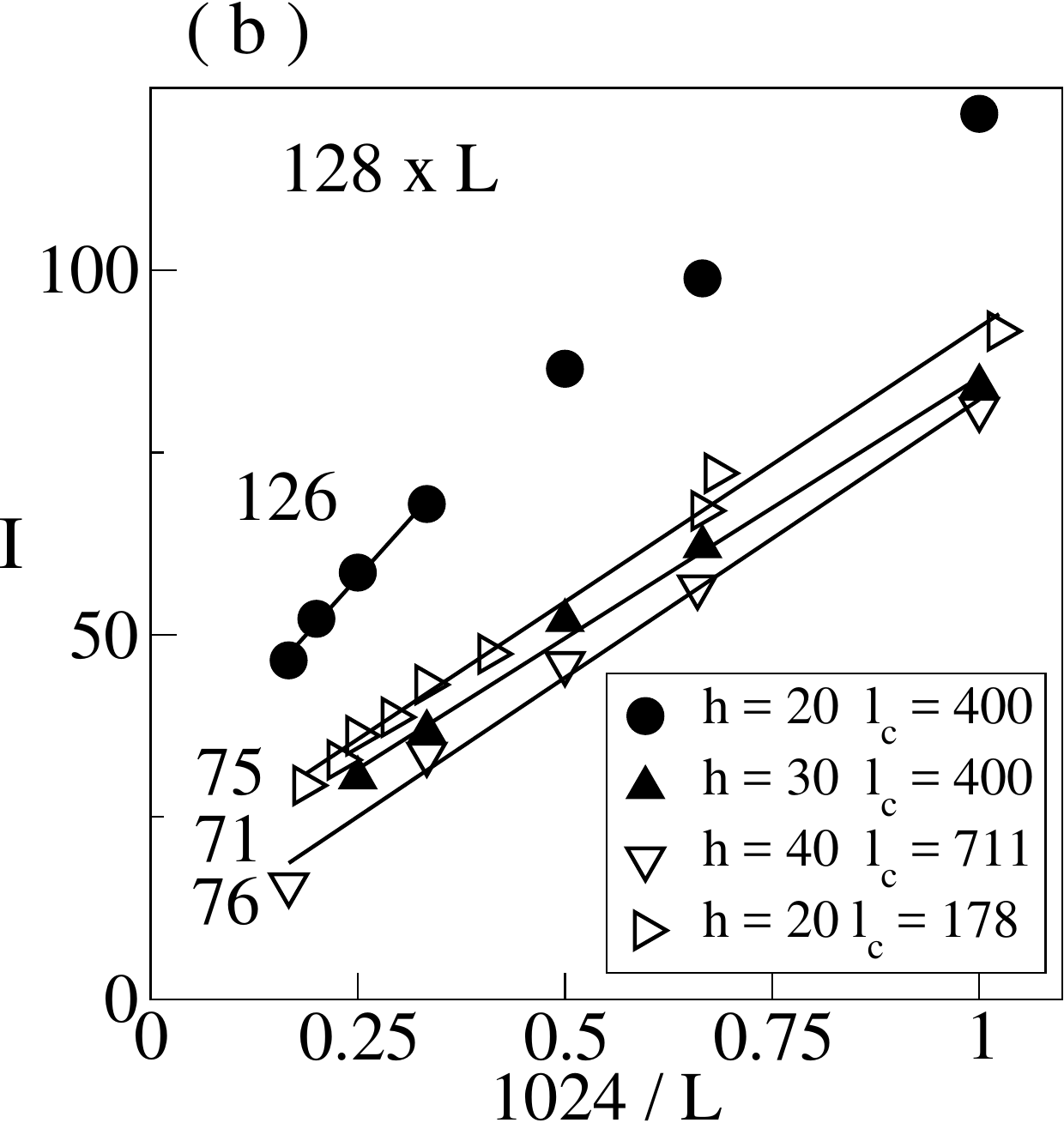}
\includegraphics[width=0.23\textwidth]{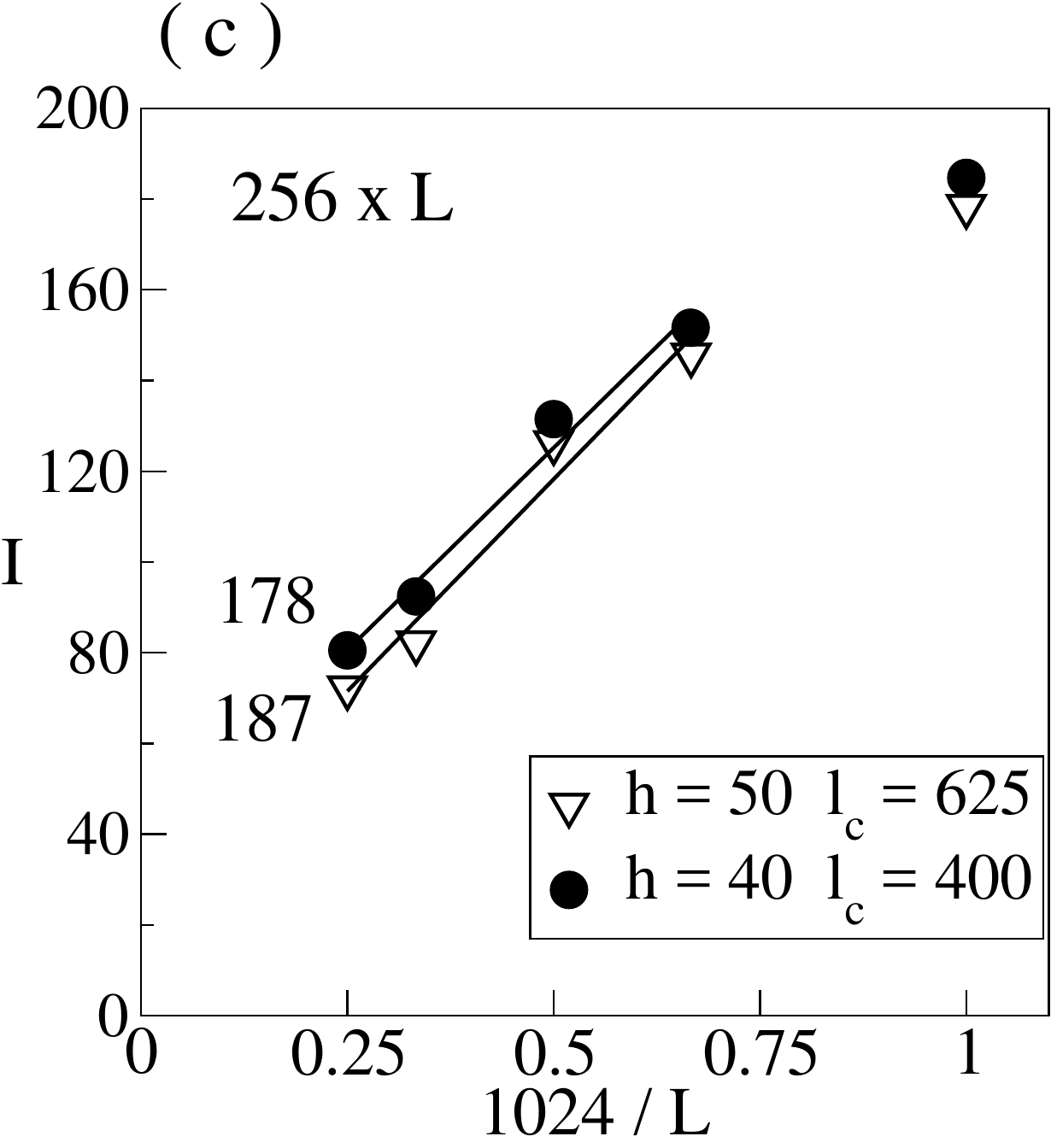}
\caption{%
The length dependence of the conductivity for various values of $\xi$ and three widths of the samples $d = 96$ (a),  128  (b) and 256 (c). Note that the  slope of linear width (given in legend)  is proportional to $\xi = \sqrt{l_c}/h$.
}
\label{fig-ADL}
\end{figure}

To prove that the universal linear dependence $I(\xi)\propto \xi$ corresponds to diffusive transport, we show in 
Fig. \ref{fig-ADL} the length-dependence of $I$ for various values of $d$ and $\xi$. 
In the diffusive regime we find $I(\xi,L)\sim \xi/L$.
For instance,  
 in Fig.  \ref{fig-ADL}(a)  the slope increases from 49  for $\xi = 0.5$ to 101 for $\xi = 1$.
In Fig. \ref{fig-ADL}(b) the slope increases from 75 to 126 when $\xi$ increases from 2/3 to 1. 
Similarly, Figs. \ref{fig-ADL}(b) and (c) confirm that samples with the same value of $\xi$ 
possess the same slope (within the accuracy of numerical data). 
Finally, obtained values of the slope agree, at least qualitatively, with predicted $d^{3/2}$-dependence on the width of the sample.  For example comparing $I(d_1=96)$ at $\xi=1/2$ (solid triangles in Fig.~\ref{fig-ADL}(a)) with $I(d_2=256)$ for same $\xi=1/2$ (open triangles in Fig.~\ref{fig-ADL}(c)) at a fixed length, e.g. $1024/L=0.25$, one can check that $I(d_1)/I(d_2)\approx 20/80=0.25$ agrees with $(d_1/d_2)^{3/2}=0.23$ within the numerical accuracy.

To summarize, we conclude that the diffusive phonon transport is universal and
\begin{equation}
I \propto \displaystyle{\sqrt{\frac{l_c}{d}}}\frac{d}{h}\frac{d}{L} \propto d^{3/2}.
\end{equation}
This result differs from the diffusive regime in 2D samples with bulk disorder, where the conductance $g$ is given by the equation 
\cite{pichard}
\be
g = \frac{\ell N}{L}.
\ee
Here $\ell$ is a mean free path of coherent scattering and $N\propto d$ is a number of open channels. 
Thus, we expect that for bulk disorder, $I\propto d$, in contrast to $I\sim d^{3/2}$ dependence for samples with surface disorder. It should be emphasized that the $d^{3/2}$ dependence of the thermal conductance $K$ corresponds to a $d^{1/2}$ dependence of the thermal conductivity $\kappa$, which depends on the single parameter $\zeta$. We note that our result can be expressed in terms of an effective mean-free path that depends not only on the geometric parameter \cite{MFP1} $d$, but also on the surface disorder parameters \cite{MFP2,WS4} $l_c$ and $h$. We emphasize that our model is valid only in the surface-roughness dominated regime.

\section{Discussion and summary}

In our numerical simulations we considered transmission of phonons across wires with surface corrugation. On the other hand our theoretical model has localized phonons, characterized by phenomenological parameters, that arise from a mapping of the surface-disordered wire to a smooth-wire with additional pseudo-interactions \cite{ma}.  It is therefore not clear how the phenomenological parameters of the localized phonons are related to the parameters characterizing the surface roughness.  We can think of the localized phonon modes  as corresponding to the resonances created inside the corrugations in our numerical simulations. 
In Eq.~ (\ref{scaling}), we used qualitative arguments to argue that  $\Gamma$  must increase with increasing $l_c$ and $d$, but decrease with increasing $h$. Our  numerical simulation suggests that $\Gamma$ can be related to a simple power-law combination of all the parameters $h$, $l_c$ and $d$, given by $\Gamma \propto l_c^{1/2}d^{3/2}/h$, consistent with the above expectations. This immediately predicts a $d^{3/2}$ dependence of the thermal conductance and therefore a $d^{1/2}$ dependence of $\kappa(T)$. We note that the parameters $h$ and $l_c$ appear in the roughness power spectrum in the combination $l_c^{1/2}/h$, which is also the same combination that appears in $\Gamma$, and therefore in the final result for $\kappa(T)$.

The breakdown of universality is clearly seen in Fig.~\ref{fig-ADD} for either large $l_c$ or small $h$, both cases leading to a ballistic regime. The small $d/h$ regime on the other hand corresponds to the localized regime where the universality also fails. While theory calculates thermal conductivity, numerical simulations obtain thermal conductance as a function of length $L$. In general, there is a non-universal part independent of $L$ coming from e.g. the contact resistance, and the universality holds only when the non-universal part is small compared to the diffusive part. As our results show, the diffusive regime, with a clear $1/L$ dependence of the thermal conductance, follows the universal behavior.

As predicted theoretically and confirmed numerically, the final result for $\kappa$ as a function of $T$ shows that in the surface-roughness dominated regime, the low-$T$ behavior is always $T^2$, and the high-$T$ behavior is independent of $T$. Both of these are clearly violated in case of the VLS wires \cite{li}. On the other hand for the ELE wires the high-$T$ behavior is clearly satisfied as noted before, and the low-$T$ behavior is consistent with figures in Ref.~[\onlinecite{hochbaum}]. The prediction of the $d^{1/2}$ behavior is clearly violated in case of VLS wires with diameters $d=37$, $56$ and $115$ nm., while it is again  consistent with figures in Ref.~[\onlinecite{hochbaum}] with similar range of diameters $d=50$, $98$ and $115$ nm. Figure 3(a) of Ref.~[\onlinecite{lim}], with three different values of the three parameters ($h$, $l_c$, $d$; $\zeta$)=(4.3, 8.4, 69.7 nm; 5.6), (2.7, 8.4, 79.8 nm; 9.6) and (2.3, 8.9, 77.5 nm; 11.4)  all of which seem to be in the surface disorder dominated regime, is consistent with the single parameter description of our model. Thus we argue that while the VLS wires are in the bulk-disorder dominated regime, the ELE wires are in the surface-roughness dominated regime. It then implies that similarly produced ELE wires should follow the universal one-parameter behavior, allowing flexibility in the choice of the various geometrical and disorder parameters to keep $\kappa$ fixed at a low value.

\section*{Acknowledgements}
We acknowledge financial support by the Slovak Research and Development Agency under the contract No. APVV-15-0496
and by the Agency  VEGA under the contract No. 1/0108/17. KAM thanks Department of Experimental Physics, Faculty of  Mathematics, Physics  and Informatics at Comenius University for kind hospitality during his visit in Fall, 2017, when this work was partially carried out.

\end{document}